\documentclass[a4paper,showpacs,showkeys,preprintnumbers,aps,prl]
{revtex4}

\usepackage{amsmath}
\usepackage{amssymb}
\usepackage{revsymb}
\usepackage{graphicx}
\usepackage{dcolumn}

\begin{document}

\title{Covariant Baryon Charge Radii and Magnetic Moments in a\\
Chiral Constituent Quark Model
\footnote{This work was supported by the Austrian Science Fund, project
no. P14806-TPH.}}
\author{K. Berger, R. F. Wagenbrunn, W. Plessas}
\affiliation{\mbox{Theoretische Physik,
Karl-Franzens-Universit\"at, Graz, Austria}}

\begin{abstract}
The charge radii and magnetic moments of all the light and strange baryons
are investigated within the framework of a constituent quark model based
on Goldstone-boson-exchange dynamics. Following the point-form approach to
relativistic quantum mechanics, the calculations are performed in a
manifestly covariant manner. Relativistic (boost) effects have a 
sizeable influence on the results. The direct predictions of the
constituent quark model are found to fall remarkably close to the
available experimental data.
\end{abstract}


\maketitle

\section{Introduction}
The charge radii and magnetic moments, as measures for the 
distribution of charge and magnetisation, represent important 
observables in hadronic physics. While quantum chromodynamics (QCD) is 
accepted as the fundamental theory of strong interactions, one is 
still lacking its direct predictions for this kind of observables. The 
reason is simply that QCD cannot (yet) be solved - accurately enough - 
in the low-energy domain of hadron ground-state properties, even 
though one has recently made progress in deducing hadron masses and 
also has obtained first results for hadron charge radii and magnetic
moments from lattice QCD \cite{AFT:02,GHPRS:01,Jones:02} (for a review
see, e.g., \cite{Bali:02}).

In view of the difficulties of solving QCD at low energies one has 
developed effective (field) theories and/or effective models. A 
promising approach is offered by constituent quark models (CQMs).
Modern CQMs can be constructed so as to include the
relevant properties of
QCD in the low-energy regime, notably the consequences of the
spontaneous breaking of chiral symmetry (SB$\chi$S). At the same time 
the requirements of special relativity can be incorporated by making 
the theory Poincar\'e-invariant.

Some time ago the Graz group suggested an interesting new type of CQM 
\cite{GPVW:98}. Its dynamics was motivated by the idea that at low 
energies the dominant QCD degrees of freedom are furnished by 
constituent quarks and Goldstone bosons \cite{GlozRisk:96}. The 
so-called Goldstone-boson-exchange (GBE) CQM for baryons relies on 
the Hamilton operator \cite{GPVW:98}
\begin{equation}
\hat{H} = \sum_{i=1}^{3}\sqrt{\hat{\vec{p}}_i\,^2+\hat{m_i}^2} +
\sum_{i<j=1}^{3} \left[\hat{V}_{conf}(ij)+ \hat{V}_{hf}(ij) \right] \,
\label{Ham}
\end{equation}
which is defined on the $\{QQQ\}$ Hilbert space $\mathcal{H}$.
The first term represents the relativistic kinetic-energy operator, 
where $\hat{\vec{p}}_i$ and $\hat{m}_i$ are the three-momentum and mass
operators of the constituent quarks, respectively. The confinement potential
$\hat{V}_{conf}(ij)$ is taken in linear form with a strength in
accordance with the string tension of QCD. The hyperfine interaction
$\hat{V}_{hf}(ij)$ consists of the spin-spin part of the exchange 
of octet and singlet pseudoscalar bosons (mesons) and produces a 
specific spin-flavor dependence \cite{GPVW:98,GPPVW:98}. With this 
hyperfine interaction the level
orderings of positive- and negative-parity states in the $N$ and 
$\Lambda$ spectra can be reproduced simultaneously in agreement with 
experiment (the detailed spectra of the GBE CQM may be found in
refs.~\cite{GPVW:98,GPPVW:98}). The Hamiltonian of eq.~(\ref{Ham}) also 
lends itself to an invariant mass operator in relativistic 
Hamiltonian dynamics (RHD) \cite{KP:91}. This turns out to be very
important for obtaining relativistic results of all kind of observables
in hadron reactions.

Beyond spectroscopy the GBE CQM has in the meantime been put to 
testing its performance with regard to the electroweak structure of 
the nucleons. Specifically one has produced its predictions for the 
nucleon elastic electromagnetic \cite{WBKPR:01} and axial form factors
\cite{GRWBKP:01}. Working in the framework of the point-form version 
of RHD \cite{Dirac:49,Klink:98} one has obtained covariant results for 
all elastic observables relating to nucleon electroweak form factors 
\cite{BGKPRW:01}. The direct predictions of the GBE CQM have been 
found to be in close agreement with all existing data in the
low-momentum transfer domain (up to a few GeV$^{2}$).

In this work we present first covariant results of the GBE CQM for
charge radii and magnetic moments of all the octet and decuplet
baryon ground states. We again apply the point-form version of RHD.
Whenever possible the theoretical predictions are compared 
to experiments. 
\section{Formalism of the Point Form}
The point form was already defined in 1949 by Dirac \cite{Dirac:49} when
he studied RHD with regard to the (smallest) stability groups of Poincar\'e 
generators of an interacting system. It is characterized by the fact 
that only the four-momentum operator $\hat{P}^{\mu}$ is affected by
interactions. All other generators of the Poincar\'e group remain 
interaction-free. As a result the spatial rotations and, most 
importantly, the Lorentz boosts are purely kinematic. Consequently,
the theory is manifestly covariant.

In the point form the problem of solving the Hamiltonian of eq.~(\ref{Ham})
is completely equivalent to the solution of the eigenvalue
problem of the mass operator $\hat{M}=\hat{M}_{free}+\hat{M}_{int}$, where
the interaction $\hat{M}_{int}$ is added to the free mass operator
$\hat{M}_{free}=\sqrt{\hat{P}_{free}^{\mu}\hat{P}_{free,\mu}}$ according to
the Bakamjian-Thomas (BT) construction \cite{BT:53}. Correspondingly the
four-momentum operator gets split into a free part $\hat{P}_{free}^{\mu}$ and
an interaction part $\hat{P}_{int}^{\mu}$
\begin{equation}
\hat{P}^{\mu} = \hat{P}_{free}^{\mu}+\hat{P}_{int}^{\mu} =
\hat{M} \hat{V}^{\mu} = (\hat{M}_{free}+\hat{M}_{int})\hat{V}^{\mu}.
\end{equation}
It is a peculiarity of the point form that in the BT construction
the four-velocity remains interaction-free:
$\hat{V}^{\mu}=\hat{V}_{free}^{\mu}$. 
When solving the eigenvalue equation of the interacting mass operator
\begin{equation}
\hat{M}|\Psi_{B}\rangle = M_B|\Psi_{B}\rangle
\end{equation}
one obtains the eigenvalues $M_B$ and the eigenstates $|\Psi_{B}\rangle$ 
for baryon $B$ on the Hilbert space $\mathcal{H}$. Due to the 
commutation relations of the Poincar\'e algebra the $|\Psi_{B}\rangle$ 
are simultaneous eigenstates of $\hat{P}^{\mu}$ ($\mu=0,1,2,3$) and 
thus also of the Hamiltonian $\hat{H}$ of eq.~(\ref{Ham}). Furthermore,
they are simultaneously eigenstates of the four-velocity operator
$\hat{V}^{\mu}$ and evidently of the total-angular-momentum 
operator $\hat{J}$ and its z-component $\hat{\Sigma}$. Therefore we
shall subsequently characterize them by the corresponding eigenvalues 
and express them as $|v_B,M_B,J,\Sigma\rangle$.\\
The $\{QQQ\}$ Hilbert space $\mathcal{H}$ is spanned by the free states
\begin{equation}
\label{3bstates}
|p_1,\sigma_1;p_2,\sigma_2;p_3,\sigma_3\rangle = |p_1,\sigma_1\rangle \otimes
|p_2,\sigma_2\rangle \otimes |p_3,\sigma_3\rangle \, ,
\end{equation}
which are direct products of free single-particle states 
$|p_i,\sigma_i\rangle$, with $p_{i}$ and $\sigma_i$ denoting the 
individual (free) four-momenta and spin projections, respectively.  
In point form, instead of working with the usual three-body states in 
eq.~(\ref{3bstates}), one introduces so-called velocity states. They 
can be constructed by applying a specific Lorentz boost
$B(v)$ to the free three-body states
$|k_1,\mu_1;k_2,\mu_2;k_3,\mu_3\rangle$ in the centre-of-momentum 
frame (for which $\sum_i \vec{k}_i=0$): 
\begin{equation}
|v;\vec{k}_1,\mu_1;\vec{k}_2,\mu_2;\vec{k}_3,\mu_3\rangle =
U_{B(v)}|k_1,\mu_1;k_2,\mu_2;k_3,\mu_3\rangle =\prod_{i=1}^{3}\sum_{\sigma_i}
D_{\sigma_i\mu_i}^{\frac{1}{2}}[R_W(k_i,B(v))]|p_1,\sigma_1;p_2,\sigma_2;
p_3,\sigma_3\rangle \, . 
\end{equation}
These velocity states also span the whole Hilbert space $\mathcal{H}$.
They have the important advantage that under general Lorentz 
transformations the occurring Wigner $D$-functions are the same for 
all three particles and the individual momenta are all rotated by the
same amount (what is not the case for the three-particle states of
eq.~(\ref{3bstates})). Of course, the practical calculations are 
facilitated a lot by expressing the baryon mass eigenstates
$|\Psi_{B}\rangle$ in the velocity-state representation
\begin{equation}
\langle v;\vec{k}_1,\mu_1;\vec{k}_2,\mu_2;\vec{k}_3,\mu_3|
v_B,M_B,J,\Sigma \rangle \sim \delta^3 (\vec{v} - \vec{v}_B)
\Psi_{M_B J \Sigma}(\vec{k}_1,\mu_1;\vec{k}_2,\mu_2;\vec{k}_3,\mu_3) \, ,
\label{eq:velcomp}
\end{equation}
where $\vec{v}$ and $\vec{v}_B$ are the total three-velocities of the
bra and ket states, respectively.
\subsection{Invariant Nucleon Form Factors}
For the elastic electromagnetic form factors one has to compute
matrix elements of the current operator $\hat{J}^{\mu}(x)$ between the
baryon states. The electromagnetic current operator itself is an
irreducible tensor operator of the Poincar\'e group. One can apply a
generalized Wigner-Eckart theorem and decompose the matrix elements into
Clebsch-Gordan coefficients times reduced matrix elements, which are the
invariant form factors \cite{Klink:98}.
Due to the covariance properties of the matrix elements one can use any
reference frame and proceed with the calculation. We choose the usual Breit
frame, where the Clebsch-Gordan coefficients are unity.
The elastic invariant form factors are then given by the matrix elements of the
current operator $\hat{\mathcal{J}}^{\mu}(0)$ between incoming and outgoing
baryon states boosted to the (standard) Breit frame
\begin{equation}
\label{emffbreit}
2M_B F^{\mu}_{\Sigma'\Sigma}(Q^2) = \langle v'_{B},M_B,J,\Sigma'|
\hat{\mathcal{J}}^{\mu}(0)|v_{B},M_B,J,\Sigma \rangle .
\end{equation}
The invariant momentum transfer $q^2 = -Q^2$ along the $z$-axis is defined
as the difference between the final and the initial four-momenta of the
baryon, $P^{\mu}\,'=M_B v'_B$ and $P^{\mu}=M_B v_B$, respectively,
\begin{equation}
q^{\mu}=(0,0,0,Q)= P^{\mu}\,'-P^{\mu} \, ,
\end{equation}
where $v_{B}$ and $v'_{B}$ are the initial and final baryon four-velocities.

The invariant form factors are related to the electric and magnetic
Sachs form factors $G_E$ and $G_M$, respectively. In the Breit frame
one has for spin-$\frac{1}{2}$ baryons
\begin{equation}
\label{sachs1}\textstyle
G_E = F^{\mu=0}_{\frac{1}{2}\frac{1}{2}} , \qquad 
G_M = \frac{2M_B}{Q} F^{\mu=1}_{\frac{1}{2} -\frac{1}{2}} 
\end{equation}
and for spin-$\frac{3}{2}$ baryons \cite{DDM:62,NW:72}
\begin{equation}
\label{sachs2}
G_E = \frac{1}{2}\left( F^{\mu=0}_{\frac{1}{2}\frac{1}{2}} +
F^{\mu=0}_{\frac{3}{2}\frac{3}{2}} \right) , \qquad
G_M = \frac{6}{5}\frac{M_B}{Q} \left( F^{\mu=1}_{\frac{1}{2}-\frac{1}{2}} +
\sqrt{3} F^{\mu=1}_{\frac{3}{2}\frac{1}{2}} \right).
\end{equation}
\subsection{Point Form Current Model}

In case of a relativistic three-body system, the matrix elements in
eq.~(\ref{emffbreit}) cannot be calculated with 
the full structure of the electromagnetic current operator. Rather
one has to resort to simplifications. For the three-body electromagnetic
current operator we therefore assume a spectator model,
i.e. only a single quark directly couples to the virtual 
photon while the other two act as spectators
\begin{eqnarray}
\label{mecurr}
\lefteqn{\hspace{-0.3cm}\langle p'_1,\sigma'_1;p'_2,\sigma'_2;p'_3,\sigma'_3
|\hat{\mathcal{J}}^{\mu}(0)| p_1,\sigma_1;p_2,\sigma_2;p_3,\sigma_3
\rangle}   \nonumber\\
& = &
3 \left(\frac{M_B^2}{M_{free}M'_{free}}\right)^{\frac{3}{2}}
\delta^3(\vec{p}\,{}'_2-\vec{p}_2)\delta^3(\vec{p}\,{}'_3-\vec{p}_3)
\times \delta_{\sigma_2'\sigma_2} \delta_{\sigma_3'\sigma_3}\, 2E_2 2E_3
\,\langle p'_1,\sigma'_1|\hat{\mathcal{J}}^{\mu}_{[1]}(0)|p_1,\sigma_1\rangle
\, , 
\end{eqnarray}
where $E_i=\sqrt{\vec{p}_i\,{}^2+m_i^2}$ ($i=2,3$) are the energies
of the spectator quarks. The single-particle current is taken in the usual 
form for a pointlike Dirac particle with charge $e_{n}$ ($n=1,2,3$)
\begin{equation}
\label{singleme}
\langle p'_n,\sigma'_n|\hat{\mathcal{J}}^{\mu}_{[1]}(0)|p_n,\sigma_n\rangle =
e_{n} \, \bar{u}(p'_n,\sigma'_n) \gamma^{\mu} u(p_n,\sigma_n) \, ,
\end{equation}
where the quark spinor can be expressed in terms of the two-component
Pauli spinor $\chi$ in the following way
\begin{equation}
u(p,\sigma)= \sqrt{E+m}\left(\begin{array}{c}
\chi \\
\frac{\vec{\sigma}\cdot\vec{p}}{E+m}\chi
\end{array}\right) \, .
\end{equation}
Eqs.~(\ref{mecurr}) and (\ref{singleme}) define the so-called point-form
spectator approximation (PFSA).
It should be noted that the PFSA current model does not represent a 
pure one-body operator. Even though the virtual photon couples only 
to a single quark, also the spectator quarks participate in the 
process. The whole baryon experiences a boost due to the total
momentum transfer $q^2$, which is then shared by all quarks. Only
a part of it is transferred to the struck quark, namely,
\begin{equation}
\tilde{q}\,{}^2 = (p'_n-p_n)^2 = [B(v'_{B})k'_n - B(v_{B})k_n]^2 .
\end{equation} 
Furthermore, since in point form the momenta are affected by interactions,
the PFSA current rather represents an effective dynamical three-body
current.

In PFSA, the matrix elements of the current operator are finally given
by the multiple integral
\begin{eqnarray}
\label{emffpfsa}
\lefteqn{F^{\mu}_{\Sigma'\Sigma}(Q^2) = 3 \int
d\vec{k}_2 d\vec{k}_3 d\vec{k}'_2 d\vec{k}'_3 
\left(\frac{M_B^2}{M_{free}M'_{free}}\right)^{\frac{3}{2}}
\sqrt{\frac{\omega_2'\omega_3'}{\omega_2\omega_3}}
\delta^3[k'_2-B^{-1}(v'_{B})B(v_{B})k_2]} \nonumber\\
&\times& \delta^3[k'_3-B^{-1}(v'_{B})B(v_{B})k_3]\,\Psi^{\ast}_{M_{B}J\Sigma'}
(\vec{k}'_1,\mu'_1;\vec{k}'_2,\mu'_2;\vec{k}'_3,\mu'_3)\,{}
{D^{1/2}_{\sigma'_1\mu'_1}}^{\ast}[R_W(k'_1,B(v'_{B}))] \nonumber\\
&\times& \frac{1}{2\sqrt{\omega_1\omega_1'}}
\,{}\langle p'_1,\sigma'_1|\hat{\mathcal{J}}^{\mu}_{[1]}(0)|p_1,\sigma_1\rangle
\,{}D^{1/2}_{\sigma_1\mu_1}[R_W(k_1,B(v_{B}))]
\Psi_{M_{B}J\Sigma}(\vec{k}_1,\mu_1; \vec{k}_2,\mu_2; \vec{k}_3,\mu_3)
\nonumber \\
&\times& D^{1/2}_{\mu'_2\mu_2}[R_W(k_2,B^{-1}(v'_{B})B(v_{B}))]\,{} 
D^{1/2}_{\mu'_3\mu_3}[R_W(k_3,B^{-1}(v'_{B})B(v_{B}))],
\end{eqnarray}
where $\omega_n=\sqrt{\vec{k}_n\,^2+m_n^2}$ are the single-quark 
energies in the baryon centre-of-momentum frame.

\subsection{Charge Radii and Magnetic Moments}
The baryon charge radii $r_{ch}^2$ and magnetic moments $\mu$ can be 
calculated from the electric and magnetic Sachs form factors of
eqs.~(\ref{sachs1}) - (\ref{sachs2}) in the limit $Q^2\to 0$:
\begin{eqnarray}                                                        
r_{ch}^2 \equiv -6\,{}\left. \frac{dG_E}{d(Q^2)} \right|_{Q^2=0} ,\qquad
\mu \equiv G_M(Q^2=0) \, .                                           
\end{eqnarray}                                                          

\section{Results and Discussion}

The predictions of the GBE CQM for the charge radii and magnetic moments
are given in Tables \ref{chexp} and \ref{magexp} for all octet and
decuplet baryons, where experimental data are available. The quoted PFSA 
results are manifestly covariant and they are immediately found
in good or reasonable agreement
with experiment in all cases. Their quality is similar to the one of 
the nucleon electroweak form factors obtained before
\cite{WBKPR:01,GRWBKP:01,BGKPRW:01}. This is remarkable because the 
results represent direct predictions obtained just with the quark 
model eigenstates, without introducing any further phenomenological 
parametrisations such as quark form factors etc.

We have also studied the influence of relativity on the results. 
First we compare to the calculations 
in nonrelativistic impulse approximation (NRIA), see Tables 
\ref{chcomp} and \ref{magcomp}.
For the charge radii the shortcomings of the nonrelativistic approach 
are immediately evident. Considerable effects are caused by both the 
relativistic current operator (cf. columns 1 and 2 of 
Table \ref{chcomp}) and the relativistic boosts (cf. column 3 of
Table \ref{chcomp}). Only the covariant results (last column of
Table \ref{chcomp}) turn out to be reasonable and compare well with 
experimental data, whenever such a comparison is possible (see Table 
\ref{chexp}).

For the magnetic moments seemingly good results are obtained with
the NRIA, especially in case of the nucleons and some other octet 
baryons. However, this has to be considered as accidental. Already
in the decuplet baryons, the influence of relativity becomes rather
large. From the relativistic (PFSA)
calculation, which employs a relativistic current operator and 
also includes boost effects, one learns 
that both of these ingredients are necessary in order to produce a
reasonable prediction in concordance with experiment (cf. the last 
three columns in Table \ref{magcomp}). In other words: In the covariant 
calculation, relativistic effects (both from the current and the 
boosts) appear even though the magnetic moments are observables at 
$Q^{2}=0$. In case a nonrelativistic current is used, Lorentz boosts 
have essentially no effect, since they enter only in higher orders of
$Q$. As a consequence the deviations from the covariant results may
become considerable (up to about 30\%, for example, in the case of 
$\Delta^{+}$).

In summary, it appears evident that, in order to reach consistent 
results, any calculation of both the charge radii and the
magnetic moments must be performed in a fully relativistic manner,
employing a relativistic current operator as well as including Lorentz
boosts. This is true even though we deal here with observables in the
limit $Q^{2} \to 0$. The point-form approach appears as a reasonable 
framework to obtain covariant predictions for charge radii as well as
magnetic moments.

\addcontentsline{toc}{chapter}{Bibliography}

\begin{thebibliography}{10}

\bibitem{AFT:02}
C. Alexandrou, P. de~Forcrand, and A. Tsapalis, Phys. Rev. D {\bf 66},  094503
  (2002).

\bibitem{GHPRS:01}
M. Goeckeler {\it et~al.}, arXiv:hep-ph/010810  (2001).

\bibitem{Jones:02}
M.~K. Jones, Nucl. Phys. A {\bf 699},  124c  (2002).

\bibitem{Bali:02}
G.~S. Bali, {\em in: {NSTAR 2002}, ({Proceedings} of the {Workshop} on the
  {Physics} of {Excited} {Nucleons}, {Pittsburgh})} (World Scientific,
  Singapore, 2003).

\bibitem{GPVW:98}
L.~Y. Glozman, W. Plessas, K. Varga, and R.~F. Wagenbrunn, Phys. Rev. D {\bf
  58},  094030  (1998).

\bibitem{GlozRisk:96}
L.~Y. Glozman and D.~O. Riska, Phys. Rep. {\bf 268},  263  (1996).

\bibitem{GPPVW:98}
L.~Y. Glozman {\it et~al.}, Phys. Rev. C {\bf 57},  3406  (1998).

\bibitem{KP:91}
B.~D. Keister and W.~N. Polyzou, Adv. Nucl. Phys. {\bf 20},  225  (1991).

\bibitem{WBKPR:01}
R.~F. Wagenbrunn {\it et~al.}, Phys. Lett. B {\bf 511},  33  (2001).

\bibitem{GRWBKP:01}
L.~Y. Glozman {\it et~al.}, Phys. Lett. B {\bf 516},  183  (2001).

\bibitem{Dirac:49}
P.~A.~M. Dirac, Rev. Mod. Phys. {\bf 21},  392  (1949).

\bibitem{Klink:98}
W.~H. Klink, Phys. Rev. C {\bf 58},  3587  (1998).

\bibitem{BGKPRW:01}
S. Boffi {\it et~al.}, Eur. Phys. J. A {\bf 14},  17  (2002).

\bibitem{BT:53}
B. Bakamjian and L.~H. Thomas, Phys. Rev. {\bf 92},  1300  (1953).

\bibitem{DDM:62}
I. L.~Durand, P. DeCelles, and R. Marr, Phys. Rev. {\bf 126},  1882  (1962).

\bibitem{NW:72}
N.~R. Nath and H.~J. Weber, Phys. Rev. D {\bf 6},  1975  (1972).

\bibitem{PDG:02}
K. Hagiwara {\it et~al.}, Phys. Rev. D {\bf 66},  010001  (2002).

\bibitem{Kotulla:02}
M. Kotulla {\it et~al.}, Acta Phys. Polon. B {\bf 33},  957  (2002).

\end{thebibliography}

\begin{table}[hb]
\caption{PFSA predictions of the GBE CQM for baryon charge radii 
$r_{ch}^2$ [fm$^2$] in comparison to experiment \cite{PDG:02}.}
\label{chexp}\begin{ruledtabular}
\begin{tabular}{lrc}
Baryon & \multicolumn{1}{r}{GBE CQM \hspace*{-1.4em}} & Experimental \\
       & \multicolumn{1}{r}{PFSA}    & Data\\
\hline
$p$          & $0.82$ & $0.757\pm0.014$\\
$n$          & $-0.13$ & $-0.1161\pm0.0022$\\
$\Sigma^{-}$ & $0.72$ & $0.61\pm0.12\pm0.09$\\
\end{tabular}
\end{ruledtabular}
\end{table}

\begin{table}[hb]
\caption{PFSA predictions of the GBE CQM for baryon magnetic moments
(in n.m.) compared to experiment \cite{PDG:02}.}
\label{magexp}\begin{ruledtabular}
\begin{tabular}{lrc}
Baryon & \multicolumn{1}{r}{GBE CQM \hspace*{-1.4em}} & Experimental \\
       & \multicolumn{1}{r}{PFSA}    & Data \\
\hline
$p$ & $2.70$  & $2.7928$\\
$n$ & $-1.70$ & $-1.9130$\\
$\Lambda$ & $-0.65$ & $-0.613\pm0.004$\\
$\Sigma^{+}$ & $2.35$  & $2.458\pm0.010$\\
$\Sigma^{-}$ & $-0.92$ & $-1.160\pm0.025$\\
$\Xi^{0}$ & $-1.24$ & $-1.250\pm0.014$\\
$\Xi^{-}$ & $-0.68$ & $-0.6507\pm0.0025$\\
$\Delta^{+}$  & $2.08$  & $2.7^{+1.0}_{-1.3}\pm1.5\pm3$\footnote{This
result is taken from ref.~\cite{Kotulla:02}.}\\
$\Delta^{++}$ & $4.17$  & $3.7 - 7.5$\\
$\Omega^{-}$ & $-1.59$ & $-2.020\pm0.05$\\
\end{tabular}
\end{ruledtabular}
\end{table}

\begin{table}[ht]
\caption{Results for charge radii (in fm$^2$) of all octet and
decuplet baryons in NRIA, a calculation using the relativistic current 
and no boosts, a calculation using the nonrelativistic current with 
boosts included, and in PFSA.} \label{chcomp}
\begin{ruledtabular}
\begin{tabular}{lrrrr}
 Baryon    & NRIA & RC w/o & NRC +    & PFSA \\
           &      & Boosts & Boosts   & \\
\hline
$p$ & $0.10$  & $0.18$  & $0.58$  & $0.82$\\
$n$ & $-0.01$ & $-0.01$ & $-0.09$ & $-0.13$\\
$\Lambda$ & $0.01$ & $0.02$ & $0.01$ & $0.03$\\
$\Sigma^{0}$ & $0.02$ & $0.03$ & $0.14$ & $0.20$\\
$\Sigma^{+}$ & $0.12$ & $0.21$ & $0.77$ & $1.13$\\
$\Sigma^{-}$ & $0.09$ & $0.16$ & $0.50$ & $0.72$\\
$\Xi^{0}$ & $0.01$ & $0.03$ & $-0.17$ &  $-0.19$\\
$\Xi^{-}$ & $0.10$ & $0.16$ & $0.41$  &  $0.54$\\
$\Sigma^{*0}$ & $0.02$ & $0.03$ & $0.01$ & $0.03$\\
$\Sigma^{*+}$ & $0.20$ & $0.32$ & $0.32$ & $0.42$\\
$\Sigma^{*-}$ & $0.16$ & $0.25$ & $0.29$ & $0.37$\\
$\Delta^{+}$  & $0.15$ & $0.25$ & $0.32$ & $0.43$\\
$\Delta^{0}$  & $0.00$ & $0.00$ & $0.00$ & $0.00$\\
$\Delta^{++}$ & $0.15$ & $0.25$ & $0.32$ & $0.43$\\
$\Delta^{-}$  & $0.15$ & $0.25$ & $0.32$ & $0.43$\\
$\Xi^{*0}$ & $0.04$ & $0.07$ & $0.03$ & $0.06$\\
$\Xi^{*-}$ & $0.16$ & $0.24$ & $0.26$ & $0.33$\\
$\Omega^{-}$ & $0.16$ & $0.22$ & $0.24$ & $0.29$\\
\end{tabular}
\end{ruledtabular}
\end{table}

\begin{table}[h]
\caption{Results for magnetic moments (in n.m.) of all octet and
decuplet baryons in NRIA, a calculation using the relativistic current 
and no boosts, a calculation using the nonrelativistic current with 
boosts included, and in PFSA.} \label{magcomp}
\begin{ruledtabular}
\begin{tabular}{lrrrr}
 Baryon      & NRIA & RC w/o & NRC +   & PFSA\\
             &      & Boosts & Boosts  & \\
\hline
$p$ & $2.74$  & $1.31$  & $2.74$  & $2.70$\\
$n$ & $-1.82$ & $-0.85$ & $-1.82$ & $-1.70$\\
$\Lambda$ & $-0.61$ & $-0.37$ & $-0.61$ & $-0.65$\\
$\Sigma^{0}$ & $0.81$  & $0.39$  & $0.81$  & $0.72$\\
$\Sigma^{+}$ & $2.63$  & $1.25$  & $2.62$  & $2.35$\\
$\Sigma^{-}$ & $-1.01$ & $-0.46$ & $-1.01$ & $-0.92$\\
$\Xi^{0}$ & $-1.40$ & $-0.79$ & $-1.40$ & $-1.24$\\
$\Xi^{-}$ & $-0.53$ & $-0.38$ & $-0.53$ & $-0.68$\\
$\Sigma^{*0}$ & $0.29$  & $0.08$  & $0.29$  & $0.09$\\
$\Sigma^{*+}$ & $3.06$  & $1.62$  & $3.06$  & $2.07$\\
$\Sigma^{*-}$ & $-2.47$ & $-1.45$ & $-2.47$ & $-1.89$\\
$\Delta^{+}$  & $2.76$  & $1.51$  & $2.76$  & $2.08$\\
$\Delta^{0}$  & $0.00$  & $0.00$  & $0.00$  & $0.00$\\
$\Delta^{++}$ & $5.52$  & $3.03$  & $5.52$  & $4.17$\\
$\Delta^{-}$  & $-2.76$ & $-1.51$ & $-2.76$ & $-2.08$\\
$\Xi^{*0}$ & $0.59$  & $0.17$  & $0.59$  & $0.18$\\
$\Xi^{*-}$ & $-2.17$ & $-1.38$ & $-2.17$ & $-1.73$\\
$\Omega^{-}$ & $-1.88$ & $-1.29$ & $-1.88$ & $-1.59$\\
\end{tabular}
\end{ruledtabular}
\end{table}

\end{document}